\documentclass[conference]{IEEEtran}
\usepackage{cite}

\usepackage[pdftex]{graphicx}
\graphicspath{{figures/}}
\DeclareGraphicsExtensions{.pdf,.jpeg,.png}

\usepackage{amsmath}
\usepackage{algorithmic}
\usepackage{array}

\usepackage{dblfloatfix}

\usepackage{url}

\usepackage{bbm}
\usepackage{algorithm}
\usepackage{xcolor}
\usepackage{caption}
\usepackage{subcaption}

\IEEEoverridecommandlockouts

\begin{document}

\title{Digital Twin-Based Multiple Access Optimization and Monitoring via Model-Driven Bayesian Learning}

\author{
    \IEEEauthorblockN{
        Clement Ruah,\IEEEmembership{~Student Member,~IEEE,}
        Osvaldo Simeone,\IEEEmembership{~Fellow,~IEEE,} and
        Bashir Al-Hashimi\IEEEmembership{,~Fellow,~IEEE}
    }
    \IEEEauthorblockA{
        Department of Engineering, King’s College London, London, UK
    }
    \thanks{
        C. Ruah and O. Simeone are with King’s Communications, Learning \& Information Processing (KCLIP) Lab.
        The work of O. Simeone was supported by the European Research Council (ERC) under the European Union's Horizon 2020 Research and Innovation Programme (grant agreement No. 725732) and by an Open Fellowship of the EPSRC.
        The work of C. Ruah was supported by the Faculty of Natural, Mathematical, and Engineering Sciences at King's College London.
    }
}

% The paper headers
% \markboth{Draft}%
% {Draft}

\IEEEtitleabstractindextext{%
    \begin{abstract}
    Commonly adopted in the manufacturing and aerospace sectors, digital twin (DT) platforms are increasingly seen as a promising paradigm to control and monitor software-based, ``open'', communication systems, which play the role of the physical twin (PT).
    In the general framework presented in this work, the DT builds a Bayesian model of the communication system, which is leveraged to enable core DT functionalities such as control via multi-agent reinforcement learning (MARL) and monitoring of the PT for anomaly detection.
    We specifically investigate the application of the proposed framework to a simple case-study system encompassing multiple sensing devices that report to a common receiver.
    The Bayesian model trained at the DT has the key advantage of capturing epistemic uncertainty regarding the communication system, e.g., regarding current traffic conditions, which arise from limited PT-to-DT data transfer.
    Experimental results validate the effectiveness of the proposed Bayesian framework as compared to standard frequentist model-based solutions.
    \end{abstract}
    
}

\maketitle

% To allow for easy dual compilation without having to reenter the
% abstract/keywords data, the \IEEEtitleabstractindextext text will
% not be used in maketitle, but will appear (i.e., to be "transported")
% here as \IEEEdisplaynontitleabstractindextext when the compsoc 
% or transmag modes are not selected <OR> if conference mode is selected 
% - because all conference papers position the abstract like regular
% papers do.
\IEEEdisplaynontitleabstractindextext
% \IEEEdisplaynontitleabstractindextext has no effect when using
% compsoc or transmag under a non-conference mode.

% For peer review papers, you can put extra information on the cover
% page as needed:
% \ifCLASSOPTIONpeerreview
% \begin{center} \bfseries EDICS Category: 3-BBND \end{center}
% \fi
%
% For peerreview papers, this IEEEtran command inserts a page break and
% creates the second title. It will be ignored for other modes.
\IEEEpeerreviewmaketitle

\section{Introduction}

\subsection{Context and Motivation} \label{subsec:context_and_motivation}

\begin{figure}[t]
    \centering
    \includegraphics[keepaspectratio, width=8.5cm]{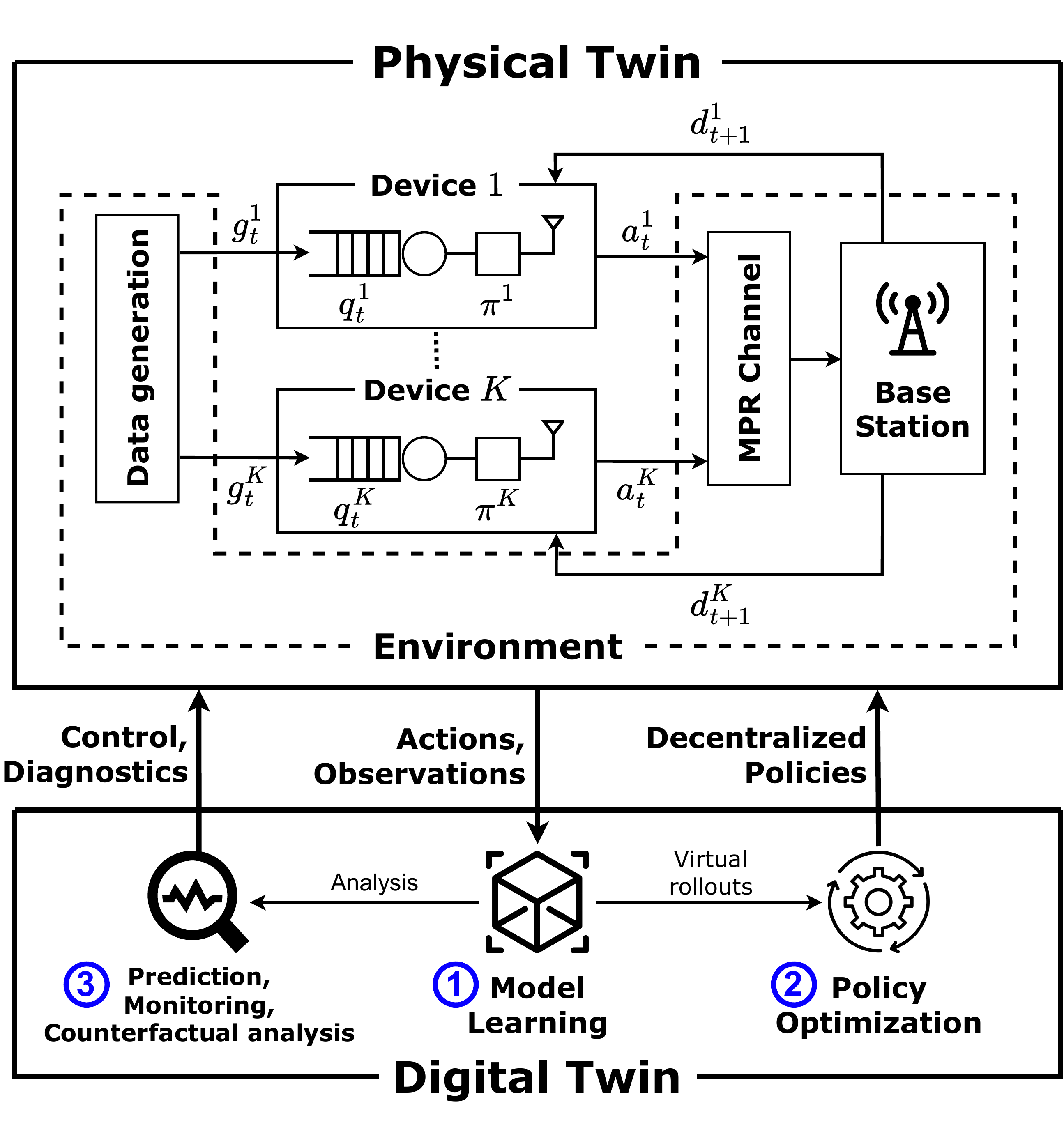}
    \vspace{-0.4cm}
    \caption{
    The digital twin (DT) platform for the control and analysis of the communication system studied in this work.
    The physical twin (PT) consists of a group of $K$ devices receiving correlated data and communicating over a shared multi-packet reception (MPR) channel.
    The DT platform operates along the phases of model learning \textcircled{\raisebox{-0.9pt}{1}} and policy optimization \textcircled{\raisebox{-0.9pt}{2}}; while also enabling functionalities such as prediction, counterfactual analysis and monitoring \textcircled{\raisebox{-0.9pt}{3}}.
    }
    \vspace{-0.3cm}
    \label{fig:DT_workflow_MAC}
\end{figure}

\IEEEPARstart{A}{} digital twin (DT) platform can be viewed as a cyberphysical system in which a physical entity, referred to as the physical twin (PT), and a virtual model, known as the DT, interact based on an automatized bi-directional flow of information \cite{grieves2017digital, kritzinger2018digital}.
Based on data received from the PT, the DT maintains an up-to-date model of the PT \cite{glaessgen2012digital}, which is leveraged to control, monitor, and analyze the operation of the PT \cite{almasan2022network}.
DT platforms are increasingly regarded as an enabling technology for wireless cellular systems built on the open networking principles of disaggregation and virtualization \cite{akman2020oran}, which are expected to be central to 6G \cite{khan2022digital}.

This paper introduces a general framework based on Bayesian learning for the design of a DT platform implementing the functions of control and monitoring of a wireless system (see Fig. \ref{fig:DT_workflow_MAC}).
In the proposed framework, the DT builds a \emph{Bayesian} model of the system dynamics based on data received from the PT.
The model is then leveraged to optimize transmission policies via Bayesian \emph{multi-agent reinforcement learning} (MARL), while also enabling monitoring for anomaly detection, without the need to integrate additional data from the PT.
Unlike standard frequentist models, Bayesian models have the key property of providing an accurate quantification of epistemic uncertainty arising from limited PT-to-DT communication, and hence limited data at the DT.
As a possible embodiment of the proposed approach, the DT platform may be implemented as an xApp, or as a collection of connected xApps, that run in the near-real-time RAN Intelligent Controller (RIC) of an Open-RAN (O-RAN) architecture \cite{lacava2022programmable}.

As an exemplifying case study, we consider a multi-access PT system consisting of a radio access network (RAN) similar to that studied in \cite{kassab2020multi, valcarce2021toward, miuccio2022learning}.
It is emphasized that, unlike \cite{kassab2020multi, valcarce2021toward, miuccio2022learning}, our goal here is not to address a particular task via MARL, but rather to introduce a general framework supporting the implementation of multiple functionalities at the DT, including control via MARL and monitoring, despite the limited data transfer from the PT to the DT.

\subsection{Related Work}

DT-aided control of PT systems is often formulated as a model-based reinforcement learning (RL) problem in which the DT is leveraged as simulation platform to optimize the PT policy \cite{dai2020deep, cronrath2019enhancing, wang2022digital}.
% \textcolor{blue}{
% While model-based methods enable safe exploration and optimization of new policies, a common shortcoming is that errors in the model generally lead to sub-optimal policies when applied by the PT in the real environment.
% A solution to this problem is presented in \cite{cronrath2019enhancing}, where a sub-optimal policy obtained from the DT is used to safely explore the real environment and train a more refined policy through model-free RL.
% Conversely, the DT can be used to explore alternative actions in between model-free training steps at the PT \cite{wang2022digital}.
% }
In \cite{li2017dynamic}, a Bayesian model is deployed by the DT to enable the quantification epistemic uncertainty.
Existing work on DT platforms for wireless systems has investigated mechanisms for DT-PT synchronization \cite{hashash2022edge} and DT-aided network optimization and monitoring \cite{hui2022digital}, as well as DT-based control for computation offloading via model-based RL \cite{dai2020deep}.
To the best of our knowledge, applications of DT relying on Bayesian learning for communication systems have not been reported in the literature.

% \textcolor{red}{
% When the PT system consists of multiple separate agents, the control problem can be framed within a MARL framework.
% A common solution implements \emph{centralized training with decentralized execution} (CTDE) \cite{kraemer2016multi}, whereby training is done in a centralized fashion, while policies at each agent rely only on local observations.
% CTDE algorithms can be implemented using value-based methods, often relying on value-decomposition networks \cite{sunehag2017value}; using actor-critic methods, often relying on the \emph{centralized critic with decentralized actors} (CCDA) paradigm \cite{lowe2017multi, foerster2018counterfactual}; or combining both methods \cite{wang2020off}.
% }
% \textcolor{red}{
% Application of MARL in telecommunications include the optimization of multi-access signaling protocols \cite{miuccio2022learning}, dynamic spectrum access \cite{kassab2020multi} and network routing \cite{tao2001multi}.
% }

\subsection{Main Contributions}

The main contributions of this paper are as follows:

\noindent $\bullet$ We propose a Bayesian framework to control and monitor an AI-native wireless system, using a multi-access RAN as an exemplifying case study \cite{kassab2020multi, valcarce2021toward, miuccio2022learning}.
In the proposed approach, as illustrated in Fig. \ref{fig:DT_workflow_MAC}, a Bayesian model learning phase is followed by policy optimization and monitoring phases, which leverage the uncertainty quantification capacity of Bayesian models.

\noindent $\bullet$ A key challenge in the definition of a Bayesian model is the choice of a \emph{domain-specific} factorization of the joint distribution of all variables of interest. This paper elucidates this design choice for a multi-access system consisting of sensing devices with correlated packet arrivals reporting to a common receiver through a shared multi-packet reception (MPR) channel \cite{tong2001multipacket}. This case study is relevant for Internet-of-Things (IoT) and machine-type communications.

\noindent $\bullet$ Experimental results confirm the advantages of the proposed Bayesian framework as compared to conventional frequentist model-based approaches in terms of metrics such as throughput and buffer overflow, as well as area under the receiver operating curve (ROC) for anomaly detection at the DT.

% \textcolor{blue}{
% The rest of the paper is organized as follows.
% Section \ref{sec:DT_control_monitoring} presents the proposed Bayesian framework for model learning, policy optimization, and anomaly detection.
% Section \ref{sec:case_study} presents the studied multi-access system and its DT platform using the proposed framework.
% In section \ref{sec:experiments}, we carry an experiment to test the proposed solution against oracle-based and frequentist model-based benchmarks.
% Finally, Section \ref{sec:conclusion} concludes the paper.
% }

\section{Bayesian DT Framework} \label{sec:DT_control_monitoring}

In this section, we formally define a PT system comprising multiple network elements, such as mobile devices or infrastructure nodes, referred to as \emph{agents}.
We then present a Bayesian DT framework to estimate the PT dynamics, optimize the agents decisions, and monitor possible anomalies in the PT.

\subsection{Multi-Agent PT System}

The PT system of interest consists of $K$ agents, e.g., $K$ sensing devices, indexed by $k \in \mathcal{K}=\{1, \dots, K\}$ that operate over a discrete time index $t = 1, 2, \dots$, e.g., over time slots.
At each time $t$, each agent takes an \emph{action} $a^k_t$, e.g., a decision on whether to transmit a packet from its queue or not.
The action is selected by following a \emph{policy} that leverages information collected by the agent regarding the current \emph{state} $s_t$ of the overall system, which may include include, e.g., packet queue lengths.
The state $s_t$ evolves according to some \emph{ground-truth transition probability} $T(s_{t+1} | s_t, a_t)$, such that the probability distribution of the next state $s_{t+1} \sim T(s_{t+1} | s_t, a_t)$ depends on the current state $s_t$ and joint action $a_t = ( a^1_t, \dots, a^K_t )$.
We restrict our framework to the case of \emph{jointly observable} states \cite{oliehoek2016concise}, in which the state $s_t$ can be identified if one has access to all \emph{observations} $o^k_t$ made by all agents $k \in \mathcal{K}$ at time $t$, i.e., in which the state is a function of the collection $o_t = ( o^1_t, \dots, o^K_t )$ for all times $t$.

Agents in the PT cannot communicate, and hence the overall information available at agent $k$ up to time $t$ is contained in its \emph{action-observation history} $h^k_t = ( o^k_1, a^k_1, o^k_2, \dots, a^k_{t-1}, o^k_t )$.
Accordingly, the behaviour of agent $k$ is defined by its \emph{policy} $\pi^k(a^k_t | h^k_t)$, which defines the probability of each possible action $a^k_t$ based on the available information $h^k_t$.

\begin{figure}[t]
    \centering
    \includegraphics[keepaspectratio, width=8.5cm]{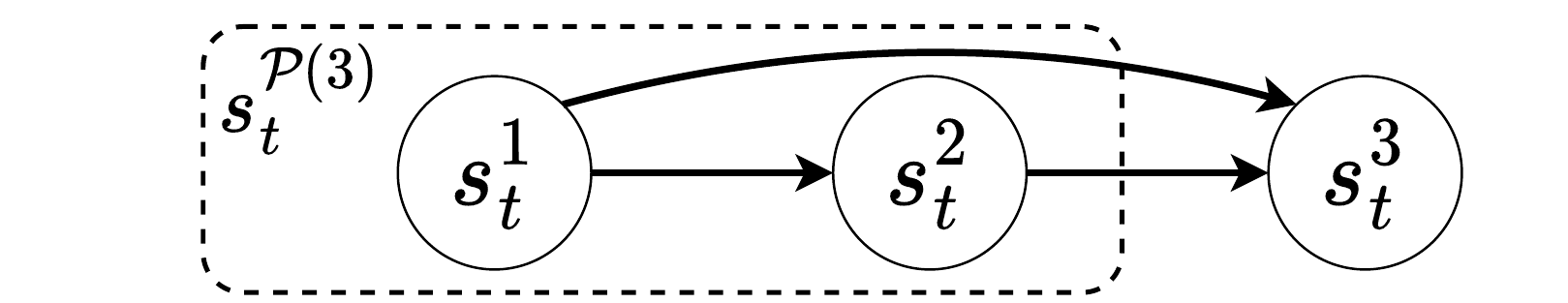}
    \vspace{-0.1cm}
    \caption{
    Example of factorization in (\ref{eq:factored_system_dynamics}), excluding the variables corresponding to previous time steps $t-1$.
    }
    \vspace{-0.4cm}
    \label{fig:DAG_example}
\end{figure}

The state of a communication system typically comprises several substates, describing, e.g., the current traffic conditions or the quality of the wireless channel.
As a result, one can typically partition the state variables $s_t$ into $M$ operationally distinct subsets $\{ s^i_t \}_{i=1}^{M}$ indexed by $i$.
To describe the interactions among these subsets, we introduce a Bayesian network defined by a directed acyclic graph, as illustrated in Fig. \ref{fig:DAG_example}, in which each subset $s^i_t$ is directly affected by the subset of ``parent'' variables $s^{\mathcal{P}(i)}_t \subseteq s_t$ (see, e.g., \cite{koller2009probabilistic}).
Accordingly, the transition probability is assumed to factorize as
\begin{equation}
\label{eq:factored_system_dynamics}
    T\left( s_{t+1} | s_t, a_t \right) = \prod_{i=1}^{M} T^i \left( s^i_{t+1} \middle| s^{\mathcal{P}(i)}_{t+1}, s_t, a_t \right),
\end{equation}
where the conditional distribution $T^i( s^i_{t+1} | s^{\mathcal{P}(i)}_{t+1}, s_t, a_t )$ describes the evolution of the next states variables $s^i_{t+1}$ given the current state $s_t$, action $a_t$, and parent variables $s^{\mathcal{P}(i)}_{t+1}$.
In general, the distribution $T^i( s^i_{t+1} | s^{\mathcal{P}(i)}_{t+1}, s_t, a_t )$ depends on some sufficient statistic of variables $s_t$ and $a_t$, which may be a function of subsets of such variables.
We refer to Sec. \ref{subsec:case_study_setting} for an instance of model (\ref{eq:factored_system_dynamics}).

\subsection{Model Learning} \label{subsec:framework_model_learning}

The goal of the model learning phase (phase \textcircled{\raisebox{-0.9pt}{1}} in Fig. \ref{fig:DT_workflow_MAC}) at the DT is to obtain an estimate of the PT system dynamics in (\ref{eq:factored_system_dynamics}).
To this end, the PT applies a fixed exploration policy $\pi_{e} = \{ \pi_{e}^{k}(a^k_t | h^k_t) \}_{k \in \mathcal{K}}$, and communicates to the DT the experiences $\mathcal{D}^{\pi_e}_{T} = \{ (s_{\tau}, a_{\tau}, s_{\tau+1}) \}_{\tau < T}$ collected up to some time $t = T$ while following policy $\pi_e$.

Using data $\mathcal{D}^{\pi_e}_{T}$, the DT optimizes parametric models $T^i_{\theta^i}( s^i_{t+1} | s^{\mathcal{P}(i)}_{t+1}, s_t, a_t )$ for each of the ground-truth distributions $T^i( s^i_{t+1} | s^{\mathcal{P}(i)}_{t+1}, s_t, a_t )$ for $i \in \{1, \dots, M\}$ in (\ref{eq:factored_system_dynamics}), yielding the overall model $T_{\theta}( s_{t+1} | s_t, a_t ) = \prod_{i=1}^{M} T^i_{\theta^i}( s^i_{t+1} | s^{\mathcal{P}(i)}_{t+1}, s_t, a_t )$.
Some of the distributions $T^i_{\theta^i}( s^i_{t+1} | s^{\mathcal{P}(i)}_{t+1}, s_t, a_t )$ may be known, e.g., those describing the operation of a communication protocol.
For such distributions, the DT simply sets $T^i( s^i_{t+1} | s^{\mathcal{P}(i)}_{t+1}, s_t, a_t ) = T^i_{\theta^i}( s^i_{t+1} | s^{\mathcal{P}(i)}_{t+1}, s_t, a_t )$.
For the unknown factors, the DT adopts a \emph{Bayesian learning} framework, which aims at evaluating the posterior distributions $P(\theta^i | \mathcal{D}^{\pi_e}_{T})$ of each model parameter vector $\theta^i$ given the available data $\mathcal{D}^{\pi_e}_{T}$.

If the observations $\{ o^k_t \}_{k \in \mathcal{K}}$, and hence the state $s_t$, take discrete values, the parameters $\theta^i$ may be chosen to directly represent the transition distributions.
In such case, if the state space is sufficiently small, the exact computation of the posterior can be done using the Dirichlet-Categorical model (see, e.g., \cite{simeone2022machine}).
For high dimensional and continuous problems, computing the exact posterior $P(\theta^i | \mathcal{D}^{\pi_e}_{T})$ is intractable, and the DT must fall back to using function approximation methods such as Bayesian neural networks (BNN), and inference algorithms such as Markov chain Monte Carlo (MCMC), or variational inference \cite{simeone2022machine}.
By (\ref{eq:factored_system_dynamics}), and assuming that the parameters $\theta^i$ are a priori independent, this comes with no loss of optimality.

\subsection{Policy Optimization} \label{subsec:framework_policy_optimization}

Having obtained the posterior distributions $P(\theta^i | \mathcal{D}^{\pi_e}_T)$ for the model parameters, during policy optimization (phase \textcircled{\raisebox{-0.9pt}{2}} in Fig. \ref{fig:DT_workflow_MAC}), the DT aims to optimize the \emph{decentralized policy} $\pi = \{ \pi^k(a^k_t | h^k_t) \}_{k \in \mathcal{K}}$ so as to maximize some user-specified performance criterion.
This is defined by a \emph{reward} function $r(s_t, a_t, s_{t+1})$, which determines in turn the \emph{total discounted return}
\begin{equation}
\label{eq:total_discounted_return}
    G_t = \sum_{\tau=t}^{+\infty} \gamma^{\tau - t} r(s_\tau, a_\tau, s_{\tau+1}),
\end{equation}
for some exponential discounting factor $\gamma \in [0, 1]$ when the PT applies the policy $\pi$.
The optimal control problem consists of the maximization of the average long-term reward
\begin{equation}
\label{eq:rl_objective}
    \max_{\pi} \mathbbm{E}_{\pi}(G_1),
\end{equation}
which amounts to a Decentralized MDP (Dec-MDP) \cite{oliehoek2016concise}.
We emphasize that the DT has only access to the model $T_{\theta}( s_{t+1} | s_t, a_t )$, and not to the ground-truth distribution in (\ref{eq:factored_system_dynamics}) when addressing problem (\ref{eq:rl_objective}).

In this work, we implement the COunterfactual Multi-Agent (COMA) algorithm in \cite{foerster2018counterfactual}, a state-of-the-art \emph{centralized critic with decentralized actors} (CCDA) method, to tackle problem (\ref{eq:rl_objective}) by leveraging model-generated virtual rollouts at the DT.
In COMA, the DT maintains a centralized critic $Q_w(s_t, a_t)$, with parameter vector $w$, as well as the decentralized policies $\pi_v = \{ \pi^k_v(a^k_t | h^k_t) \}_{k \in \mathcal{K}}$, with common parameter vector $v$.
The parameterized critic $Q_w(s_t, a_t)$ is optimized to approximate the \emph{Q-value} $Q^{\pi_v}(s_t, a_t) = \mathbbm{E}_{\pi_v}[ G_t | s_t, a_t ]$ and is used only for policy optimization.
Once optimized, each policy $\pi^k_v$ is transmitted to its respective agent $k$ at the end of the policy optimization phase.

A key distinction between the approach adopted here and the conventional COMA implementation is the fact that the model $T_{\theta}( s_{t+1} | s_t, a_t )$ is stochastic with model parameter vector $\theta$ being distributed according to the posterior $P(\theta | \mathcal{D}^{\pi_e}_{T})$.
In a manner similar to \cite{zhang2021centralized}, we account for the epistemic uncertainty encoded by the posterior $P(\theta | \mathcal{D}^{\pi_e}_{T})$ by periodically sampling a parameter vector $\theta \sim P(\theta | \mathcal{D}^{\pi_e}_{T})$ during policy optimization so as to produce the next state $s_{t+1} \sim T_{\theta}( s_{t+1} | s_t, a_t )$ in the virtual rollouts.

\subsection{Monitoring}

As an example of functionalities enabled by Bayesian model learning at the DT besides control, we focus here on anomaly detection (phase \textcircled{\raisebox{-0.9pt}{3}} in Fig. \ref{fig:DT_workflow_MAC}).
Anomaly detection aims at detecting significant changes in the dynamics of the PT.
To formulate this problem, we assume that, during the operation of the system following policy optimization (phase \textcircled{\raisebox{-0.9pt}{2}} in Fig. \ref{fig:DT_workflow_MAC}), the DT has access to the information $\mathcal{D}^{\pi}_{T^{\mathrm{M}}} = \{ (s_{\tau}, a_{\tau}, s_{\tau+1}) \}_{\tau < {T^{\mathrm{M}}}}$ about the state-action sequence experienced by the PT within some time window ${T^{\mathrm{M}}}$ under the optimized policy $\pi$.
The DT tests if the collected data $\mathcal{D}^{\pi}_{T^{\mathrm{M}}}$ is consistent with the data reported by the PT during the most recent model learning phase (phase \textcircled{\raisebox{-0.9pt}{1}} in Fig. \ref{fig:DT_workflow_MAC}), or rather if it provides evidence of changed conditions or anomalous behavior.

To this end, we adopt a \emph{disagreement-based test metric} (see, e.g., \cite{daxberger2019bayesian}), which quantifies the level of epistemic uncertainty associated with the reported experience $\mathcal{D}^{\pi}_{T^{\mathrm{M}}}$.
We define as
\begin{equation}
\label{eq:likelihood}
    LL\left( \mathcal{D}^{\pi}_{T^{\mathrm{M}}} \middle| \theta \right) = \sum_{\tau = 1}^{T^{\mathrm{M}} - 1} \log\left( T_{\theta}( s_{\tau + 1} | s_{\tau}, a_{\tau} )\pi(a_{\tau} | s_{\tau}) \right)
\end{equation}
the log-likelihood of model $\theta$ for the reported experience $\mathcal{D}^{\pi}_{T^{\mathrm{M}}}$, where $\pi(a_{\tau} | s_{\tau}) = \prod_{k \in \mathcal{K}} \pi^k(a^k_{\tau} | h^k_{\tau})$.
We then consider the test metric is given by the variance
\begin{equation}
\label{eq:monitoring_metric}
    \mathbbm{E}_{P(\theta | \mathcal{D}^{\pi_e}_{T})} \left[ \left(
        LL\left( \mathcal{D}^{\pi}_{T^{\mathrm{M}}} \middle| \theta \right) -
        \mathbbm{E}_{P(\theta | \mathcal{D}^{\pi_e}_{T})} \left[
            LL\left( \mathcal{D}^{\pi}_{T^{\mathrm{M}}} \middle| \theta \right)
        \right]
    \right)^2  \right],
\end{equation}
estimated using samples from distribution $P(\theta | \mathcal{D}^{\pi_e}_{T})$.
A larger variance provides evidence of a large epistemic uncertainty, which is taken to indicate an anomalous observation $\mathcal{D}^{\pi}_{T^{\mathrm{M}}}$ as compared to the model learning conditions.

\section{Multi-Access System Case Study} \label{sec:case_study}

In order to illustrate the operation and the benefits of the proposed framework for the implementation of a DT platform, in the rest of the paper we focus on a multi-access IoT-like wireless network as the PT system to be controlled and monitored \cite{kassab2020multi, valcarce2021toward,miuccio2022learning}.

\subsection{Setting} \label{subsec:case_study_setting}

As illustrated in Fig. \ref{fig:DT_workflow_MAC}, the PT system under study comprises $K$ sensing devices that obtain data with correlated data arrivals both in time \cite{nikaein2013simple} and across devices \cite{kassab2020multi}, and communicate with a common base station (BS) over an unknown MPR channel \cite{tong2001multipacket}.
With $t$ denoting the time slot index, and following the notation in Sec. \ref{sec:DT_control_monitoring}, each device $k \in \mathcal{K}$ observes $o^k_t = ( q^k_t, g^k_t, d^k_t )$, where $q^k_t \in \{0, 1, \dots, Q^k_{\mathrm{max}}\}$ is the number of packets in the device buffer;
$g^k_t \in \{0, 1\}$ is a binary variable indicating if a new packet is generated ($g^k_t = 1$) at time $t$ or not ($g^k_t = 0$);
and $d^k_t \in \{0, 1\}$ indicates whether a packet sent at the previous time step $t-1$ from device $k$ was successfully delivered at the BS ($d^k_t = 1$) or not ($d^k_t = 0$).
The state of the PT is represented by the joint observation of all devices, $s_t = o_t = (o^1_t, \dots, o^K_t)$.

The access policy of device $k$ is given by the distribution $\pi^k(a^k_t | h^k_t)$, where we have $a^k_t = 1$ if the device attempts to transmit the first packet in its buffer, and $a^k_t = 0$ if it stays idle in slot $t$.
Finally, we define the (binary) packet-generation vector as $g_t = (g^1_t, \dots, g^K_t)$, the successful packet-delivery vector as $d_t = ( d^1_t, \dots, d^K_t )$, and the packet-transmission vector as $a_t = ( a^1_t, \dots, a^K_t )$.

In the O-RAN implementation mentioned in Sec. \ref{subsec:context_and_motivation}, communication between the PT and the DT occurs on the E2 interface, with each device $k$ communicating its observations $o^k_t$ to the RIC, and the DT feeding back the transmission policies $\{ \pi^{k}(a^k_t | h^k_t) \}_{k \in \mathcal{K}}$ to the corresponding devices through the relevant distributed unit \cite{lacava2022programmable}.

As we will detail next, and as an instance of (\ref{eq:factored_system_dynamics}), the ground-truth transition dynamics can be factorized as
\begin{equation}
\label{eq:case_study_system_dynamics}
\begin{split}
T(s_{t+1} | s_t, a_t) = & P(g_{t+1} | g_t) \times P(d_{t+1} | a_t) \times \\
    & \prod_{k \in \mathcal{K}}  P(q^k_{t+1} | q^k_t, d^k_{t+1}, g^k_{t+1}) ,
\end{split}
\end{equation}
where $P( g_{t+1} | g_t )$ describes the packet-generation process; the distribution $P( d_{t+1} | a_t )$ describes the shared access channel; and $P( q^k_{t+1} | q^k_t, d^k_{t+1}, g^k_{t+1} )$ describes the (deterministic) buffer update rule.

The packet generation mechanism is modelled as a correlated \emph{Markovian process} with transition probability distribution $P(g_{t+1} | g_t)$.
To account for spatial correlation, we partition the devices into clusters $\{ \mathcal{C}^i \}_{i=1}^{C}$ with $\mathcal{C}^i \subseteq \mathcal{K}$, $\mathcal{C}^i \cap \mathcal{C}^j = \emptyset$ if $i \neq j$ and $\bigcup_{i=1}^{C} \mathcal{C}^i = \mathcal{K}$, where each cluster  $\mathcal{C}^i$ contains devices with correlated packet arrivals.
Accordingly, the data-generation dynamics factorize without loss of generality as
\begin{equation}
\label{eq:data_generation_distribution}
    P(g_{t+1} | g_t) = \prod_{i=1}^{C} P\left( g^{\mathcal{C}^i}_{t+1} \middle| g^{\mathcal{C}^i}_t \right),
\end{equation}
where $g^{\mathcal{C}^i}_t = \{ g^k_t \}_{k \in \mathcal{C}^i}$ for $i \in \{1, \dots, C\}$.

Each device $k$ maintains a \emph{first-in first-out buffer} of maximum capacity $Q^k_{\mathrm{max}}$, where the buffer state $q^k_t$ evolves according to the deterministic update $P(q^k_{t+1} | q^k_t, d^k_{t+1}, g^k_{t+1})$ given by
\begin{equation}
\label{eq:buffer_update_rule}
    q^k_{t+1} = \min(Q^k_{\mathrm{max}}, q^k_t + g^k_{t+1} - d^k_{t+1}).
\end{equation}
If device $k$ generates a new packet when the buffer is full and transmission fails, i.e., if we have the equalities $q^k_t = Q^k_{\mathrm{max}}$, $g^k_{t+1} = 1$, and $d^k_{t+1} = 0$, a \emph{buffer overflow} event occurs at time step $t+1$.
In this case, the oldest packet in the buffer is deleted without being sent, and the newly generated packet at time $t+1$ is added to the buffer as per the update rule in (\ref{eq:buffer_update_rule}).

The transmission probability from the devices to the BS $P(d_{t+1} | a_t)$ is described by an \emph{MPR channel} \cite{tong2001multipacket}.
Accordingly, of the $n^{\mathrm{Tx}}_t = \sum_{k \in \mathcal{K}} a^k_t \in \{1, \dots, K\}$ simultaneously sent packets at time $t$, only $n^{\mathrm{Rx}}_{t+1} = \sum_{k \in \mathcal{K}} d^k_{t+1} \in \{0, 1, \dots, n^{\mathrm{Tx}}_t\}$ are successfully received as the BS.
The distribution of the number of received packets $n^{\mathrm{Rx}}_{t+1}$ given $n^{\mathrm{Tx}}_t$ is defined by the MPR channel distribution $P(n^{\mathrm{Rx}}_{t+1} | n^{\mathrm{Tx}}_t)$. 
The $n^{\mathrm{Rx}}_{t+1}$ successfully received packets are chosen uniformly from the $n^{\mathrm{Tx}}_t$ transmissions.
Note that the action variable $a^k_t$ necessarily takes value $a^k_t = 0$ if the buffer is empty, i.e., we have the inequality $a^k_t \leq q^k_t$.
Furthermore, delivery can be successful only if a packet is transmitted, i.e., we have $a^k_t \geq d^k_{t+1}$.
Therefore, the input-output MPR channel distribution relating the packet-transmission vector $a_t$ for time $t$ to the successful packet-delivery vector $d_{t+1}$ at time $t+1$ is given by
\begin{equation}
\label{eq:mpr_channel_distribution}
    P(d_{t+1} | a_t) = P(n^{\mathrm{Rx}}_{t+1} | n^{\mathrm{Tx}}_t) \times \frac{\prod_{k \in \mathcal{K}} \mathbbm{1}_{\left\{ a^k_t \geq d^k_{t+1} \right\}}}{\binom{ n^{\mathrm{Tx}}_t }{ n^{\mathrm{Rx}}_{t+1} }}.
\end{equation}
For each successfully decoded packet, the BS sends back an acknowledgement (ACK) message to the sending device $k$ over an error-free channel on the control plane.
A summary of the notations can be found in the upmost part of Figure \ref{fig:DT_workflow_MAC}.

\subsection{Model Learning}

The deterministic queue dynamics (\ref{eq:buffer_update_rule}) and the symmetry of the MPR channel in the rightmost part of (\ref{eq:mpr_channel_distribution}) are assumed to be known to the DT, so that the DT needs only an estimate of the data generation process distribution $P( g_{t+1} | g_t )$, and of the MPR channel distribution $P( n^{\mathrm{Rx}}_{t+1} | n^{\mathrm{Tx}}_t )$.
The DT models the  $M = C+1$ factors $\{ P( g^{\mathcal{C}^i}_{t+1} | g^{\mathcal{C}^i}_t ) \}_{i=1}^{C}$ and $P( n^{\mathrm{Rx}}_{t+1} | n^{\mathrm{Tx}}_t )$ using tabular models, and we adopt the  Categorical-Dirichlet Bayesian model (see Sec. \ref{subsec:framework_model_learning}).

\subsection{Policy Optimization}

In a similar manner to \cite{miuccio2022learning}, we assume that the reward in (\ref{eq:total_discounted_return}) takes the form
\begin{equation}
    \label{eq:overall_reward_definition}
    r(s_t, a_t, s_{t+1}) = \sum_{k \in \mathcal{K}} \beta^k r^k(o^k_t, a^k_t, o^k_{t+1}),
\end{equation}
with
\begin{equation}
    \label{eq:agent_reward_definition}
    r^k(o^k_t, a^k_t, o^k_{t+1}) = \left\{
    \renewcommand{\arraystretch}{1}
    \begin{array}{ll}
        + \xi & \mbox{if } d^k_{t+1} = 1 \\
        - \xi & \mbox{if } q^k_t = Q^k_{\mathrm{max}} \mbox{, } g^k_{t+1} = 1 \mbox{ and} \\
            & d^k_{t+1} = 0 \\
        % Collision reward
        % - 1 & \mbox{if } a^k_t = 1 \mbox{ and } d^k_{t+1} = 0 \\
        %     & \mbox{(collision)} \\
        - 1 & \mbox{otherwise},
    \end{array}
    \right.
\end{equation}
where the first condition corresponds to successful packet delivery and the second condition to buffer overflow.
The constants $\{\beta^k\}_{k \in \mathcal{K}}$ and $\xi > 0$ are hyperparameters under the control of the network operator at the DT.
The average in (\ref{eq:rl_objective}) is taken with respect to the packet-generation process, the MPR channel, and the policy $\pi$, with all buffers initialized as being empty, i.e., $q^k_1 = 0$ for all devices $k \in \mathcal{K}$.

The critic $Q_w(s_t, a_t)$ and the policies $\pi^k_v(a^k_t | h^k_t)$ for the COMA algorithm presented in Sec. \ref{subsec:framework_policy_optimization} are implemented as feedforward networks.
Specifically, the policy $\pi^k_v(a^k_t | h^k_t)$ takes as input its current observation $o^k_t$, along with the positional input $p_t = (t \mod{F})$, where $F$ is a hyperparameter, resulting in a simplified policy $\pi^k(a^k_t | o^k_t, p_t)$.
The adoption of complex policies using recurrent neural networks (RNNs) \cite{zhu2017improving} is left for future work.
We refer to the codebase repository \cite{Ruah_Bayesian_Digital_Twin_2022} for further details on the implementation of policy optimization.

\section{Experiments}\label{sec:experiments}

\subsection{Setup and Benchmarks}\label{subsec:experimental_setup}

In this section, we evaluate the proposed DT platform of our case study in a simulated scenario consisting of $K=4$ devices distributed in $C=2$ clusters, with devices $1$ and $2$ in cluster $\mathcal{C}^1$, and devices $3$ and $4$ in cluster $\mathcal{C}^2$.
The per-cluster packet-generation distribution $P( g^{\mathcal{C}^i}_{t+1} | g^{\mathcal{C}^i}_t )$ in (\ref{eq:data_generation_distribution}) is such that two devices in the same cluster cannot simultaneously generate a packet, and a new packet is generated at either device with probability $0.4$.
Each device $k \in \{1, 2, 3, 4\}$ is equipped with a buffer with capacity $Q^k_{\mathrm{max}} = 1$ packet, and the reward parameters in (\ref{eq:overall_reward_definition}), (\ref{eq:agent_reward_definition}), and (\ref{eq:total_discounted_return}) are set to $\beta^k = 1$ for all $k \in \{1, 2, 3, 4\}$, $\gamma = 0.95$, and $\xi = 50$.

The MPR channel allows for the successful transmission of a single packet with probability $1$; while, for two simultaneous transmissions, one packet is received with probability 0.8 and both packets are received with probability 0.2.
More than two simultaneous transmissions cause the loss of all packets.

The Bayesian model is initialized with all prior Dirichlet parameters equal to $0.01$.
As benchmarks, we consider:
$(i)$ an \emph{oracle-aided} model-free policy optimization scheme, in which the policy optimizer is allowed to interact with the ground-truth model (\ref{eq:case_study_system_dynamics}) until convergence;
$(ii)$ a \emph{frequentist} model-based MARL approach, which obtains a maximum a posteriori (MAP) estimate $\theta^{\mathrm{MAP}}$ of the model parameter vector $\theta$ during model learning with all Dirichlet prior parameters set to $1.01$, guaranteeing well-defined solutions for the MAP problem, and uses the single optimized model $T_{\theta^{\mathrm{MAP}}}( s_{t+1} | s_t, a_t )$ for policy optimization and monitoring.
For data collection during the model learning phase, we adopt a random exploration policy $\pi_e^k( a^k_t = 1 | h^k_t ) = q_t$, where probability $q_t$ is uniformly and independently selected in the interval $[0, 1]$ at each step $t$.

\begin{figure}[t]
    \centering
    \begin{subfigure}{0.49\textwidth}
        \centering
        \includegraphics[width=\textwidth]{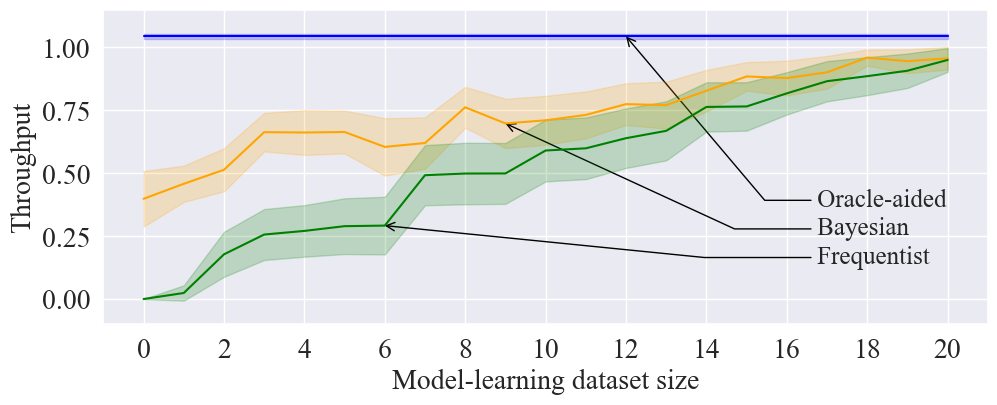}
        \vspace{-0.7cm}
        \caption{}
        \label{fig:throughput_per_model_steps}
    \end{subfigure}
    \hfill
    \begin{subfigure}{0.49\textwidth}
        \centering
        \includegraphics[width=\textwidth]{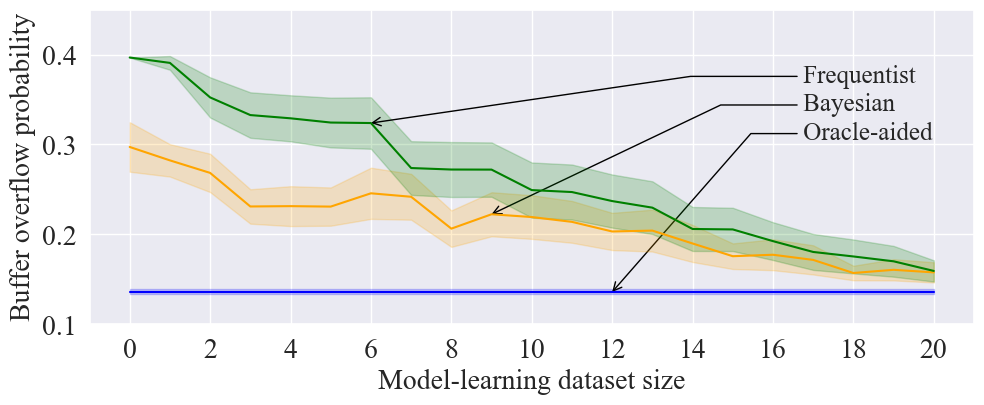}
        \vspace{-0.7cm}
        \caption{}
        \label{fig:overflow_per_model_steps}
    \end{subfigure}
    % \vspace{-0.6cm}  % For pdf
    \vspace{-0.2cm}  % For arXiv
    \caption{
    Throughput (a) and buffer overflow probability (b) as a function of the size of the dataset available in the model learning phase for the proposed Bayesian model-based approach (orange), as well as the oracle-aided model-free (blue) and frequentist model-based (green) benchmarks.
    %Metrics are averaged over time and over $50$ independent model learning and policy optimization cycles.
    }
    \vspace{-0.4cm}
    \label{fig:policies_metrics}
\end{figure}

\subsection{Policy Evaluation}

We evaluate the performance of policy optimization in the ground-truth environment by using the following metrics: the average number of packets successfully sent at each time step, i.e., the \emph{throughput} (Fig. \ref{fig:throughput_per_model_steps}); and the average probability of \emph{buffer overflow} across all devices (Fig. \ref{fig:overflow_per_model_steps}).
We focus on the impact of the size of the model learning dataset $\mathcal{D}^{\pi_e}_T$ by varying the number of exploratory steps $T$ from $0$ to $20$ in the model learning phase.
The results are averaged over 50 independent model learning and policy optimization cycles.

From Fig. \ref{fig:policies_metrics}, we observe that, in regimes with high data availability during the model learning phase, i.e., with large $T$, both Bayesian and frequentist model-based methods yield policies with similar performance to the oracle-aided benchmark.
In the low-data regime, however, Bayesian learning achieves superior performance compared to its frequentist counterpart.
With frequentist learning, which disregards epistemic uncertainty, policy optimization is prone to \emph{model exploitation}, whereby the optimized policy is misled by model errors into taking actions that are unlikely to be advantageous in the ground-truth dynamics.
By using an ensemble of models with distinct transition dynamics in state-action space regions with high epistemic uncertainty, Bayesian learning reduces the sensitivity of the optimized policy to model errors.

\subsection{Anomaly Detection}

We now consider the performance of anomaly detection by assuming that an anomalous event occurs when device $2$ is disconnected, resulting in an anomalous distribution $P( g^{\mathcal{C}^1}_{t+1} | g^{\mathcal{C}^1}_t )$ for which a packet is generated at device $1$ only with probability $0.4$, and no packet is generated at either device in cluster $\mathcal{C}^1$ with probability $0.6$.
To focus on such anomalies at the packet generation level, we use the log-likelihood $LL^{\mathcal{C}^1}( \mathcal{D}^{\pi}_{T^{\mathrm{M}}} | \theta ) = \sum_{\tau = 1}^{T^\mathrm{M} - 1} \log( P( g^{\mathcal{C}^1}_{\tau + 1} | g^{\mathcal{C}^1}_{\tau} ) )$ in place of $LL( \mathcal{D}^{\pi}_{T^{\mathrm{M}}} | \theta )$ in (\ref{eq:monitoring_metric}).
Furthermore, we consider as benchmark a standard test based on the log-likelihood $LL^{\mathcal{C}^1}( \mathcal{D}^{\pi}_{T^{\mathrm{M}}} | \theta^{\mathrm{MAP}} )$ obtained from MAP-based frequentist learning.

\begin{figure}[t]
    \centering
    \includegraphics[keepaspectratio, width=8.5cm, height=6.8cm]{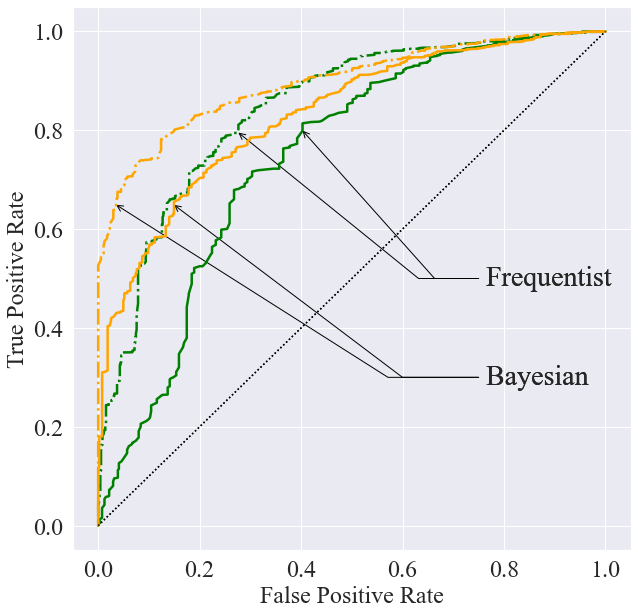}
    \vspace{-0.2cm}
    \caption{
    Receiver operating curves (ROC) of the Bayesian (orange) and frequentist (green) anomaly detection tests.
    Solid lines represent model learning dataset sizes of $T=20$ steps, while dashed lines correspond to dataset sizes of $T=50$ steps.
    }
    \vspace{-0.5cm}
    \label{fig:roc_anomaly_detection}
\end{figure}

For each model learning dataset size $T=20$ and $T=50$, we report the false positive rates (FPR) and the true positive rates (TPR) of the anomaly detection tests by varying the detection threshold in Fig. \ref{fig:roc_anomaly_detection}.
Each curve is averaged over $50$ independent model learning phases, while the optimized policy $\pi$ used to report experiences $\mathcal{D}^{\pi}_{T^{\mathrm{M}}}$ remains the same.

Bayesian anomaly detection is observed to uniformly outperform its frequentist counterpart, achieving a higher area under the ROC in Fig. \ref{fig:roc_anomaly_detection}.
For instance, at a TPR of $0.8$, the Bayesian anomaly detector has a FPR of $0.33$ for a model learning dataset size of $T=20$ and a FPR of $0.15$ for a dataset size of $T=50$; whereas the frequentist benchmark has a FPR of $0.40$ for $T=20$ and $0.28$ for $T=50$.
These results suggest that measuring epistemic uncertainty instead of likelihood can yield more effective monitoring solutions.

\vspace{-0.1cm}
\section{Conclusions} \label{sec:conclusion}

This paper has proposed a Bayesian framework for the development of a DT platform by focusing on a multi-access sensing system as a case study.
By accounting for epistemic uncertainty during model learning, the proposed Bayesian DT framework was shown to obtain more reliable control than conventional frequentist counterparts in the low data regime, while also enhancing monitoring functionalities.
Future work may investigate the use of more complex policies accounting for partial observability at each agent \cite{zhu2017improving}; and the design of exploration policies for model learning \cite{zhang2021centralized}.

\vspace{-0.1cm}
% references section
% Can use something like this to put references on a page
% by themselves when using endfloat and the captionsoff option.
\ifCLASSOPTIONcaptionsoff
  \newpage
\fi

% trigger a \newpage just before the given reference
% number - used to balance the columns on the last page
% adjust value as needed - may need to be readjusted if
% the document is modified later
%\IEEEtriggeratref{8}

\bibliographystyle{IEEEtran}
\bibliography{refs}

% Generated by IEEEtran.bst, version: 1.14 (2015/08/26)
\begin{thebibliography}{10}
\providecommand{\url}[1]{#1}
\csname url@samestyle\endcsname
\providecommand{\newblock}{\relax}
\providecommand{\bibinfo}[2]{#2}
\providecommand{\BIBentrySTDinterwordspacing}{\spaceskip=0pt\relax}
\providecommand{\BIBentryALTinterwordstretchfactor}{4}
\providecommand{\BIBentryALTinterwordspacing}{\spaceskip=\fontdimen2\font plus
\BIBentryALTinterwordstretchfactor\fontdimen3\font minus
  \fontdimen4\font\relax}
\providecommand{\BIBforeignlanguage}[2]{{%
\expandafter\ifx\csname l@#1\endcsname\relax
\typeout{** WARNING: IEEEtran.bst: No hyphenation pattern has been}%
\typeout{** loaded for the language `#1'. Using the pattern for}%
\typeout{** the default language instead.}%
\else
\language=\csname l@#1\endcsname
\fi
#2}}
\providecommand{\BIBdecl}{\relax}
\BIBdecl

\bibitem{grieves2017digital}
M.~Grieves and J.~Vickers, ``Digital twin: Mitigating unpredictable,
  undesirable emergent behavior in complex systems,'' in
  \emph{Transdisciplinary perspectives on complex systems}.\hskip 1em plus
  0.5em minus 0.4em\relax Springer, 2017, pp. 85--113.

\bibitem{kritzinger2018digital}
W.~Kritzinger, M.~Karner, G.~Traar, J.~Henjes, and W.~Sihn, ``Digital twin in
  manufacturing: A categorical literature review and classification,''
  \emph{IFAC-PapersOnLine}, vol.~51, no.~11, pp. 1016--1022, 2018.

\bibitem{glaessgen2012digital}
E.~Glaessgen and D.~Stargel, ``The digital twin paradigm for future {NASA} and
  {US} {Air} {Force} vehicles,'' in \emph{in Proc. AIAA/ASME/ASCE/AHS/ASC
  Structures, Structural Dynamics and Materials Conference 20th AIAA/ASME/AHS
  Adaptive Structures Conference 14th AIAA}, 2012, p. 1818.

\bibitem{almasan2022network}
P.~Almasan, M.~Ferriol-Galm{\'e}s, J.~Paillisse, J.~Su{\'a}rez-Varela,
  D.~Perino, D.~L{\'o}pez, A.~A.~P. Perales, P.~Harvey, L.~Ciavaglia, L.~Wong
  \emph{et~al.}, ``Network digital twin: Context, enabling technologies and
  opportunities,'' \emph{arXiv preprint arXiv:2205.14206}, 2022.

\bibitem{akman2020oran}
A.~Akman, C.~Li, L.~Ong, L.~Suciu, B.~Sahin, T.~Li, P.~Stjernholm, J.~Voigt,
  A.~Buldorini, Q.~Sun \emph{et~al.}, ``{ORAN} use cases and deployment
  scenarios: Towards open and smart {RAN},'' \emph{O-RAN Alliance, White Paper,
  Feb}, 2020.

\bibitem{khan2022digital}
L.~U. Khan, W.~Saad, D.~Niyato, Z.~Han, and C.~S. Hong, ``Digital-twin-enabled
  {6G}: Vision, architectural trends, and future directions,'' \emph{IEEE
  Communications Magazine}, vol.~60, no.~1, pp. 74--80, 2022.

\bibitem{lacava2022programmable}
\BIBentryALTinterwordspacing
A.~Lacava, M.~Polese, R.~Sivaraj, R.~Soundrarajan, B.~S. Bhati, T.~Singh,
  T.~Zugno, F.~Cuomo, and T.~Melodia, ``Programmable and customized
  intelligence for traffic steering in {5G} networks using {Open} {RAN}
  architectures,'' \emph{arXiv}, 2022. [Online]. Available:
  \url{https://arxiv.org/abs/2209.14171}
\BIBentrySTDinterwordspacing

\bibitem{kassab2020multi}
R.~Kassab, A.~Destounis, D.~Tsilimantos, and M.~Debbah, ``Multi-agent deep
  stochastic policy gradient for event based dynamic spectrum access,'' in
  \emph{2020 IEEE 31st Annual International Symposium on Personal, Indoor and
  Mobile Radio Communications}.\hskip 1em plus 0.5em minus 0.4em\relax IEEE,
  2020, pp. 1--6.

\bibitem{valcarce2021toward}
A.~Valcarce and J.~Hoydis, ``Toward joint learning of optimal {MAC} signaling
  and wireless channel access,'' \emph{IEEE Transactions on Cognitive
  Communications and Networking}, vol.~7, no.~4, pp. 1233--1243, 2021.

\bibitem{miuccio2022learning}
L.~Miuccio, S.~Riolo, S.~Samarakoon, D.~Panno, and M.~Bennis, ``Learning
  generalized wireless {MAC} communication protocols via abstraction,'' in
  \emph{GLOBECOM 2022 - 2022 IEEE Global Communications Conference}, 2022, pp.
  2322--2327.

\bibitem{dai2020deep}
Y.~Dai, K.~Zhang, S.~Maharjan, and Y.~Zhang, ``Deep reinforcement learning for
  stochastic computation offloading in digital twin networks,'' \emph{IEEE
  Transactions on Industrial Informatics}, vol.~17, no.~7, pp. 4968--4977,
  2020.

\bibitem{cronrath2019enhancing}
C.~Cronrath, A.~R. Aderiani, and B.~Lennartson, ``Enhancing digital twins
  through reinforcement learning,'' in \emph{2019 IEEE 15th International
  Conference on Automation Science and Engineering (CASE)}.\hskip 1em plus
  0.5em minus 0.4em\relax IEEE, 2019, pp. 293--298.

\bibitem{wang2022digital}
X.~Wang, L.~Ma, H.~Li, Z.~Yin, T.~Luan, and N.~Cheng, ``Digital twin-assisted
  efficient reinforcement learning for edge task scheduling,'' in \emph{2022
  IEEE 95th Vehicular Technology Conference:(VTC2022-Spring)}.\hskip 1em plus
  0.5em minus 0.4em\relax IEEE, 2022, pp. 1--5.

\bibitem{li2017dynamic}
C.~Li, S.~Mahadevan, Y.~Ling, S.~Choze, and L.~Wang, ``Dynamic {Bayesian}
  network for aircraft wing health monitoring digital twin,'' \emph{Aiaa
  Journal}, vol.~55, no.~3, pp. 930--941, 2017.

\bibitem{hashash2022edge}
O.~Hashash, C.~Chaccour, and W.~Saad, ``Edge continual learning for dynamic
  digital twins over wireless networks,'' \emph{arXiv preprint
  arXiv:2204.04795}, 2022.

\bibitem{hui2022digital}
L.~Hui, M.~Wang, L.~Zhang, L.~Lu, and Y.~Cui, ``Digital twin for networking: A
  data-driven performance modeling perspective,'' \emph{arXiv preprint
  arXiv:2206.00310}, 2022.

\bibitem{tong2001multipacket}
L.~Tong, Q.~Zhao, and G.~Mergen, ``Multipacket reception in random access
  wireless networks: From signal processing to optimal medium access control,''
  \emph{IEEE Communications Magazine}, vol.~39, no.~11, pp. 108--112, 2001.

\bibitem{oliehoek2016concise}
F.~A. Oliehoek and C.~Amato, \emph{A concise introduction to decentralized
  {POMDPs}}.\hskip 1em plus 0.5em minus 0.4em\relax Springer, 2016.

\bibitem{koller2009probabilistic}
D.~Koller and N.~Friedman, \emph{Probabilistic graphical models: principles and
  techniques}.\hskip 1em plus 0.5em minus 0.4em\relax MIT press, 2009.

\bibitem{simeone2022machine}
O.~Simeone, \emph{Machine Learning for Engineers}.\hskip 1em plus 0.5em minus
  0.4em\relax Cambridge University Press, 2022.

\bibitem{foerster2018counterfactual}
J.~Foerster, G.~Farquhar, T.~Afouras, N.~Nardelli, and S.~Whiteson,
  ``Counterfactual multi-agent policy gradients,'' in \emph{Proceedings of the
  AAAI conference on artificial intelligence}, vol.~32, no.~1, 2018.

\bibitem{zhang2021centralized}
Q.~Zhang, C.~Lu, A.~Garg, and J.~Foerster, ``Centralized model and exploration
  policy for multi-agent {RL},'' \emph{arXiv preprint arXiv:2107.06434}, 2021.

\bibitem{daxberger2019bayesian}
E.~Daxberger and J.~M. Hern{\'a}ndez-Lobato, ``Bayesian variational
  autoencoders for unsupervised out-of-distribution detection,'' \emph{arXiv
  preprint arXiv:1912.05651}, 2019.

\bibitem{nikaein2013simple}
N.~Nikaein, M.~Laner, K.~Zhou, P.~Svoboda, D.~Drajic, M.~Popovic, and S.~Krco,
  ``Simple traffic modeling framework for machine type communication,'' in
  \emph{ISWCS 2013; The Tenth International Symposium on Wireless Communication
  Systems}.\hskip 1em plus 0.5em minus 0.4em\relax VDE, 2013, pp. 1--5.

\bibitem{zhu2017improving}
P.~Zhu, X.~Li, P.~Poupart, and G.~Miao, ``On improving deep reinforcement
  learning for {POMDPs},'' \emph{arXiv preprint arXiv:1704.07978}, 2017.

\bibitem{Ruah_Bayesian_Digital_Twin_2022}
\BIBentryALTinterwordspacing
C.~Ruah, ``{Bayesian Digital Twin Repository},'' 2022. [Online]. Available:
  \url{https://github.com/kclip/bayesian-dt}
\BIBentrySTDinterwordspacing

\end{thebibliography}

% biography section

% \begin{IEEEbiography}{Clement Ruah}
% Biography text here.
% \end{IEEEbiography}

% % if you will not have a photo at all:
% \begin{IEEEbiographynophoto}{Osvaldo Simeone}
% Biography text here.
% \end{IEEEbiographynophoto}

\end{document}